\documentclass[
preprint,
superscriptaddress,
groupedaddress,
unsortedaddress,
runinaddress,
frontmatterverbose, 
preprint,
preprintnumbers,
nofootinbib,
 amsmath,amssymb,
 aps, physrev,
pra,
prb,
rmp,
prstab,
prstper,
floatfix,
]{revtex4-2}

\usepackage{graphicx}
\usepackage{dcolumn}
\usepackage{bm}
\usepackage{hyperref}

\usepackage{color}
\allowdisplaybreaks[4]

\begin{document}

\title{\textbf{Can Mirror Symmetry Challenge Local Realism? Probing Photon Entanglement from Positronium via Compton Scattering} 
}%

\author{Junle Pei}
 \email{Contact author: peijunle@hnas.ac.cn}
\affiliation{Institute of Physics, Henan Academy of Sciences, Zhengzhou 450046, P. R. China
}%

\author{Lina Wu}
 \email{Contact author: wulina@xatu.edu.cn}
\affiliation{School of Sciences, Xi'an Technological University, Xi'an 710021, P. R. China
}


\begin{abstract}
This study investigates photon entanglement generated from para-positronium decay by analyzing azimuthal correlations after the double Compton scattering with stationary electrons. We introduce a normalized correlation observable $\mathcal{O}_1 = \cos(2\phi_1 - 2\phi_2)/C_1$ to witness entanglement. In the absence of decoherence, $\langle\mathcal{O}_1\rangle = -1$, corresponding to a maximally entangled Bell state. With decoherence parameterized by $\rho$, the expectation becomes $-(1-\rho)$, allowing direct experimental quantification of coherence loss. A prior symmetry analysis of the Compton scattering process within the quantum field theory (QFT) is provided, which establishes the mirror-symmetric nature of the single-photon angular distribution. We further examine a local hidden-variable theory (LHVT) under the angular-momentum conservation and the rotation invariance. Imposing the mirror symmetry with respect to the plane defined by the photon spin and momentum leads to a non-negative LHVT prediction for $\langle \sin^2\theta_1 \sin^2\theta_2 \cos(2\phi_1-2\phi_2)\rangle$, contradicting the negative QFT prediction value for any $\rho < 1$. Thus, mirror symmetry serves as a novel criterion to exclude LHVT descriptions of the entangled state, whereas without preserving this symmetry, LHVTs can reproduce the correlations.
\end{abstract}

\maketitle


\section{Introduction} \label{sec:1} 

Quantum entanglement, a hallmark feature of quantum mechanics, has been at the forefront of foundational studies since the seminal debates between Einstein and Bohr \cite{PhysRev.47.777,PhysRev.48.696,EINSTEIN1936349,BOHR1996339}. Its nonlocal correlations challenge classical intuition and impose stringent constraints on any theory that seeks to preserve local realism—the idea that physical properties are defined prior to measurement and that influences cannot propagate faster than light. Bell inequalities \cite{PhysicsPhysiqueFizika.1.195}, derived under the assumptions of locality and realism, provide a powerful empirical criterion to test quantum mechanics against local hidden‑variable theories (LHVTs). Violations of these inequalities in experiments \cite{PhysRevLett.28.938,PhysRevLett.47.460,PhysRevLett.49.1804,Bouwmeester_1997,PhysRevLett.81.3563,PhysRevLett.80.1121,Riebe:2004jpa,Barrett:2004bxh,Hensen:2015ccp,Han:2025ewp,Afik:2020onf,Fabbrichesi:2021npl,Han:2023fci,ATLAS2024,Cheng:2023qmz,CMS:2024pts,Cheng:2025cuv,Subba:2024aut,BESIII:2018cnd,Pei:2025yvr,BESIII:2025vsr,Lin:2025eci,Wu:2025dds} have consistently favored quantum predictions, offering strong evidence against local realism.
Despite these successes, the search for new and complementary ways to probe the limits of local realism remains active \cite{ABEL1992304,Li:2024luk,Bechtle:2025ugc,Abel:2025skj,Pei:2026wfu,Pei:2026rlh}. In particular, scenarios that do not rely directly on Bell inequalities can offer fresh insights and may expose different facets of the quantum-classical divide.

Among the earliest and cleanest sources of entangled photon pairs is the two‑photon decay of para‑positronium (p‑Ps). A positron annihilating at rest with an electron produces two back‑to‑back photons in a maximally entangled polarization state. Because ideal polarization analyzers do not exist for 511 keV gamma rays, measurements of the relative linear polarization have historically been performed via Compton scattering. Wu and Shaknov \cite{PhysRev.77.136} first demonstrated good agreement with quantum mechanics \cite{Pryce1947}, and later Langhoff \cite{Langhoff1960} refined the measurement. A particularly thorough experimental and theoretical analysis was carried out by Kasday, Ullman and Wu \cite{Kasday1975} (see also Kasday’s thesis and subsequent review \cite{Kasday1971}). Using a carefully designed Compton polarimeter, they confirmed the quantum predictions for the angular correlation of scattered photons.

Crucially, Kasday et al. also examined the implications of their experiment for LHVTs. They observed that with Compton scattering alone--without additional assumptions--no direct test of Bell’s inequality is possible, because one can explicitly construct a local hidden‑variable model that reproduces all quantum predictions for the momentum distributions of the scattered photons. In fact, Bell himself provided such a counter‑example (reproduced in Appendix I of Ref.~\cite{Kasday1971}), and Kasday presented an even simpler one. The essential message of those works is: a Compton‑scattering experiment on annihilation photons, analyzed without extra assumptions, does not rule out LHVTs.
Since then, the recognized wisdom has been that only ideal, or nearly ideal, polarization measurements (e.g., in optics with high‑efficiency detectors) can close the locality loophole. However, even in optical Bell tests, auxiliary assumptions (such as fair sampling) are often required.

In recent years, Compton-polarimetry studies of p-Ps annihilation photons have explored decoherence \cite{PARASHARI2024138628}, quantum entanglement \cite{Ivashkin2023}, and nonmaximal entanglement \cite{Moskal:2024fpv}, as well as detection using plastic positron emission tomography (PET) \cite{Sharma:2023axg}, cadmium zinc telluride (CZT) PET \cite{Watts2021}, and gadolinium aluminum gallium garnet (GaGG)-based Compton polarimeters \cite{cryst10121073,Parashari_2022}.
In this work, we revisit the test of local realism from a theoretical perspective, asking a different question: Can physically motivated symmetry conditions, beyond those implicitly assumed in previous LHVT constructions, exclude local hidden‑variable descriptions of the Compton‑scattered photon pair?
Our analysis proceeds as follows. First, we derive the full joint angular distribution of the two Compton‑scattered photons for a general polarization state of the back‑to‑back photon pair from p‑Ps decay, including a phenomenological parameter $\rho$ that characterizes decoherence. This extension to partially coherent states is itself beyond the scope of the earlier studies \cite{Kasday1975,Kasday1971}, which focused on pure states.
Second, we attempt to construct an LHVT that can reproduce this distribution. Guided by angular‑momentum conservation (the decay produces a spin‑0 state), we parameterize the hidden variables as the directions of the photon spins, respecting rotation invariance. We show that without any further restriction, such an LHVT can indeed be built and it closely resembles the Bell‑type counter-examples discussed in Ref.~\cite{Kasday1971}.

The central new result of our paper is the following: when we additionally impose mirror symmetry with respect to the plane spanned by the photon momentum and its spin direction (a symmetry that holds in QED and is experimentally well established for single photon Compton scattering), the LHVT construction fails to reproduce the quantum correlation. More precisely, the expectation value $\langle \sin^2\theta_1\sin^2\theta_2\cos(2\phi_1-2\phi_2)\rangle$ becomes non‑negative under the LHVT with mirror symmetry, whereas quantum mechanics (for any $\rho<1$) predicts a negative value. This provides a new, symmetry‑based criterion to challenge local realism, one that was not discussed in the earlier works \cite{Kasday1975,Kasday1971} and is not captured by their Bell‑type counter‑examples.

Thus, the present paper does not claim to perform a better experiment or to circumvent the known “no‑go” result of Kasday et al. Instead, we offer a novel theoretical insight: the compatibility of Compton‑scattering data with local realism depends crucially on whether the hidden‑variable theory is required to respect mirror symmetry. Because mirror symmetry is implied by the parity and rotation invariance of QED and has been verified in single‑photon Compton scattering, its violation by an LHVT would render such a theory physically unnatural. Our work therefore opens a new direction for testing local realism using symmetry principles, independent of the detection‑efficiency or fair‑sampling loopholes.

The paper is organized as follows: In Section \ref{sec:2}, we introduce the entanglement witness based on Compton-scattered azimuthal correlations. Section \ref{sec:3} details the production and decoherence of photon pairs from p-Ps decay. Section \ref{sec:4} examines the feasibility of a local hidden-variable description, with emphasis on the role of mirror symmetry. We conclude in Section \ref{sec:con}.

\section{Entanglement Witness via Compton-Scattered Azimuthal Correlations} \label{sec:2} 

The polarization state of a photon pair in a pure state can be expressed as
\begin{align}  
    |\gamma_1\gamma_2\rangle=\sum_{k_1,k_2=\pm 1}\alpha_{k_1,k_2} |k_1\rangle_{\gamma_1} |k_2\rangle_{\gamma_2}~,  \label{djxs}
\end{align}  
where $ k_i~(i=1,2)$ denote the spin projection quantum numbers of photons $ \gamma_i $ along their respective momentum directions. The coefficients $ \alpha_{k_1,k_2} $ are subject to the normalization condition of
\begin{align}  
    \sum_{k_1,k_2=\pm 1}\left|\alpha_{k_1,k_2}\right|^2=1~.  \label{con1}
\end{align}  

We consider back-to-back photons $\gamma_1$ and $\gamma_2$ produced from the decay of p-Ps at rest in the laboratory frame. The photons propagate with equal magnitudes of momentum ($p_{\gamma_1}=p_{\gamma_2}=m_e$) in opposite directions. Their four-momenta are parameterized as
\begin{align}  
    & \gamma_i:(m_e,0,0,\pm m_e)~,\quad i=1,2~.\label{rsg}
\end{align}
When photons $\gamma_i$ are scattered via Compton processes with stationary electrons in the laboratory frame, i.e. $\gamma_i+e^-_i \to \gamma_i^\prime+e^{-\prime}_i$, we parameterize the four-momenta of the initial electrons $e^-_i$ and final-state photons $\gamma_i^\prime$ as
\begin{align}  
   & e_i^-:(m_e,0,0,0)~, \label{csd}\\  
   & \gamma^\prime_i:(m_eq_i ,m_eq_i\sin\theta_i\cos\phi_i,m_e q_i\sin\theta_i\sin\phi_i,m_e q_i\cos\theta_i)~,  \label{csg}
\end{align}  
where $\theta_i\in [0,\pi]$, $\phi_i \in[0,2\pi]$, and the dimensionless energy parameters $q_i$ are given by
\begin{align}  
    & q_i=\frac{1}{2\mp\cos\theta_i}~,\quad i=1,2~.  \label{q12}
\end{align}  
The four-momenta of the recoiling electrons $e^{-\prime}_i$ are determined via four-momentum conservation in the scattering processes. 
The differential phase space volume elements for final-state particles are given by
\begin{align}
    d\pi_{f_i}=\frac{1}{16\pi^2}{q_i^2}~d\cos\theta_i~d\phi_i~. 
\end{align}

We denote the scattering amplitudes for the processes $\gamma_i+e^-_i \to \gamma_i^\prime+e^{-\prime}_i~(i=1,2)$ as $\mathcal{M}^i_{k;j,m,n}$, where $k=\pm 1$, $j=\pm \frac{1}{2}$, $m=\pm 1$, and $n=\pm \frac{1}{2}$ are the spin projection quantum numbers of $\gamma_i$, $e^-_i$, $\gamma_i^\prime$, and $e^{-\prime}_i$ along their respective momentum directions. 
Direct calculations yield the scattering amplitudes
\begin{align}  
    \mathcal{M}^i_{k;j,m,n}=4\pi \alpha_{em} e^{i(\pm k+j-n)\phi_i}\tilde{\mathcal{M}}^i_{k;j,m,n}~,\quad i=1,2~,  
\end{align}  
where $\alpha_{em}$ denotes the electromagnetic fine-structure constant. The real-valued functions $\tilde{\mathcal{M}}^i_{k;j,m,n}$ exhibit angular dependence only on $\theta_i$ (not $\phi_i$), whose explicit expressions are provided in the Appendix \ref{appen}.  
For the pure $\gamma_1\gamma_2$ state described by Eq.~({\ref{djxs}}), the joint angular distribution of $\gamma_1^\prime \gamma_2^\prime$, denoted $\mathcal{W}(\theta_1,\theta_2,\phi_1,\phi_2)$, is
\begin{align}  
   \mathcal{W}(\theta_1,\theta_2,\phi_1,\phi_2)=&\frac{1}{\Gamma} \sum_{\substack{j_1,n_1 = \pm \frac{1}{2} \\ m_1=\pm1}}~\sum_{\substack{j_2,n_2=\pm \frac{1}{2}\\m_2=\pm1}} \left|\sum_{k_1,k_2=\pm 1}  \alpha_{k_1,k_2}\mathcal{M}^1_{k_1;j_1,m_1,n_1}\mathcal{M}^2_{k_2;j_2,m_2,n_2}\right|^2~.\label{ysfb}
\end{align}
To enforce the normalization condition
\begin{align}  
    \int d\pi_{f_1} \int d\pi_{f_2}~ \mathcal{W}(\theta_1,\theta_2,\phi_1,\phi_2) =1~,
\end{align}  
$\Gamma$ is defined as
\begin{align}  
   \Gamma=& \int d\pi_{f_1} \int d\pi_{f_2}~\Gamma~\mathcal{W}(\theta_1,\theta_2,\phi_1,\phi_2)  \nonumber\\
   =&4\pi^2 \alpha^4_{em} \sum_{k_1,k_2=\pm 1}\left(\left|\alpha_{k_1,k_2}\right|^2\int_{-1}^1 d\cos\theta_1~q_1^2 H^1_{k_1,k_1^\prime=k_1}   \int_{-1}^1 d\cos\theta_2 ~q_2^2 H^2_{k_2,k_2^\prime=k_2}  \right)~,
\end{align}
where
\begin{align}  
&H^i_{k,k^\prime}=\sum_{\substack{j,n = \pm \frac{1}{2} \\ m=\pm1}}\tilde{\mathcal{M}}^i_{k;j,m,n} \tilde{\mathcal{M}}^i_{k^\prime;j,m,n}~.
\end{align}
Using the expressions of $\tilde{\mathcal{M}}^i_{k;j,m,n}$ in the Appendix \ref{appen}, we obtain
\begin{align}  
&H^i_{1,1}=H^i_{-1,-1}=4\frac{2+(1\mp\cos\theta_i)^3}{2\mp\cos\theta_i}~,\quad H^i_{1,-1}=H^i_{-1,1}=4\sin^2\theta_i~,\quad i=1,2~.
\end{align}
So, we get
\begin{align}  
   \Gamma=& \alpha^4_{em} \frac{64}{81}\pi^2 \left(40-27\ln 3\right)^2~. 
\end{align}

For any angular observable $\mathcal{O}(\theta_1,\theta_2,\phi_1,\phi_2)$, the statistical average is
\begin{align}
    \langle \mathcal{O}(\theta_1,\theta_2,\phi_1,\phi_2) \rangle=& 
    \int d\pi_{f_1} \int d\pi_{f_2}~\mathcal{O}(\theta_1,\theta_2,\phi_1,\phi_2)~\mathcal{W}(\theta_1,\theta_2,\phi_1,\phi_2) \nonumber \\
    =& \sum_{k_1,k_2,k_1^\prime,k_2^\prime=\pm 1}\mathcal{O}_{k_1,k_2;k_1^\prime,k_2^\prime} \alpha_{k_1,k_2} \alpha^*_{k_1^\prime,k_2^\prime}~,
\end{align}
where
\begin{align}
    \mathcal{O}_{k_1,k_2;k_1^\prime,k_2^\prime} =&
    \left(\frac{18\pi}{40-27\ln 3}\right)^2  \int d \pi_{f_1} \int d \pi_{f_2}~ 
    \mathcal{O}(\theta_1,\theta_2,\phi_1,\phi_2)  e^{i(k_1-k_1^\prime)\phi_1} e^{-i(k_2-k_2^\prime)\phi_2} H^1_{k_1,k^\prime_1} H^2_{k_2,k^\prime_2}~.
\end{align}
For example, for the observable $\cos(2\phi_1-2\phi_2)$, we obtain
\begin{align}
    &\langle \cos(2\phi_1-2\phi_2)\rangle=C_1 \left(\alpha_{-1,-1}\alpha^*_{1,1}+\alpha_{1,1}\alpha^*_{-1,-1}\right)
\end{align}
with
\begin{align}
    C_1= 648 \frac{(1-\ln 3)^2}{(40-27\ln 3)^2}~.
\end{align}
Similarly, for fixed $\theta_1$ and $\theta_2$, the statistical average of $\cos(2\phi_1-2\phi_2)$ is
\begin{align}
    &\langle \cos(2\phi_1-2\phi_2)\rangle_{\theta_1,\theta_2}=C_1^\prime (\theta_1,\theta_2) \left(\alpha_{-1,-1}\alpha^*_{1,1}+\alpha_{1,1}\alpha^*_{-1,-1}\right)
\end{align}
with
\begin{align}
    C_1^\prime (\theta_1,\theta_2)= \frac{1}{2} \frac{(2-\cos\theta_1)\sin^2\theta_1}{2+(1-\cos\theta_1)^3} \frac{(2+\cos\theta_2)\sin^2\theta_2}{2+(1+\cos\theta_2)^3}~.
\end{align}
We introduce the normalized correlation observables  
\begin{align}  
    &\mathcal{O}_1 = \cos(2\phi_1-2\phi_2)/C_1~, \\
    &\mathcal{O}^\prime_1(\theta_1,\theta_2) =\cos(2\phi_1-2\phi_2)/C_1^\prime (\theta_1,\theta_2)~.  
\end{align}  
When expanded in terms of $\alpha_{k_1,k_2}~(k_1,k_2=\pm 1)$, both $\mathcal{O}_1$ and $\mathcal{O}^\prime_1(\theta_1,\theta_2)$ share identical expressions (denoted as $\delta$)
\begin{align}  
    \delta=\alpha_{-1,-1}\alpha^*_{1,1}+\alpha_{1,1}\alpha^*_{-1,-1}~. 
\end{align}  
The predicted range of $\delta$ under the normalization condition in Eq.~(\ref{con1}) is $[-1, 1]$. For separable states of $\gamma_1 \gamma_2$ without quantum entanglement, the coefficients can be factorized as  
\begin{align}  
    & \alpha_{k_1,k_2} = \beta_{k_1} \sigma_{k_2}~, \quad k_1,k_2 = \pm1~,  \label{gyh2}
\end{align}  
where $\beta_{k_1}$ and $\sigma_{k_2}$ describe the polarization states of $\gamma_1$ and $\gamma_2$ respectively, each satisfying its own normalization constraint of
\begin{align}  
    & \sum_{k_1 = \pm1} \left|\beta_{k_1}\right|^2 = \sum_{k_2 = \pm 1} \left|\sigma_{k_2}\right|^2 = 1~.  \label{fenjie}
\end{align}  
The predicted range of $\delta$ for separable $\gamma_1\gamma_2$ states is $[-1/2, 1/2]$. Therefore, any observed deviation from this range in either $\mathcal{O}_1$ or $\mathcal{O}^\prime_1(\theta_1, \theta_2)$ provides sufficient evidence for quantum entanglement in the $\gamma_1 \gamma_2$ system. 
This entanglement criterion generalizes straightforwardly from the pure $\gamma_1\gamma_2$ state to the mixed $\gamma_1\gamma_2$ state~\cite{Pei:2025ito}.

\section{Production and Decoherence of Photon Pairs} \label{sec:3}

This section provides a comprehensive treatment of the generation and possible degradation of photon-pair entanglement from p-Ps decay. We begin by deriving the explicit form of the two-photon quantum state resulting from the annihilation of spin-singlet p-Ps, establishing its maximally entangled Bell-state character. We then introduce a joint angular distribution that encodes the polarization correlations after both photons undergo Compton scattering, and we examine how decoherence, modeled as a partial loss of quantum coherence, modifies these distributions. The analysis allows us to connect measurable azimuthal correlations directly to the degree of entanglement and to quantify decoherence effects in a experimentally accessible way.

\subsection{Photon Pair Production from Para-Positronium Decay}\label{sec:3-1} 

Since the QED scattering matrix is crossing symmetric, the p-Ps decay and Compton scattering are related by crossing.
At the leading order (LO) level ($v=0$) in the the non-relativistic limit of QED, we can consider p-Ps as a bound state of $e^-e^+$. The momentum-space quantum state of p-Ps can be expressed as
\begin{align}
 |\text{p-Ps}\rangle=&\frac{\sqrt{m_\text{p-Ps}}}{2m_e} \int \frac{d^3 \vec{k}}{(2 \pi)^3} \widetilde{\psi}(\vec{k}) \left(\left|\lambda_{e^-}=\frac{1}{2};\vec{k}\right\rangle_{e^-}\left|\lambda_{e^+}=\frac{1}{2};-\vec{k}\right\rangle_{e^+}\right. \nonumber\\
&\left. +\left|\lambda_{e^-}=-\frac{1}{2};\vec{k}\right\rangle_{e^-}\left|\lambda_{e^+}=-\frac{1}{2};-\vec{k}\right\rangle_{e^+}\right)~, \label{ppss}
\end{align}
where $\widetilde{\psi}(\vec{k})$ represents the wave function in momentum space, $\vec{k}$ ($-\vec{k}$) is the three-momentum of $e^-$ ($e^+$) in the rest frame of p-Ps satisfying the condition $\left|\vec{k}\right|\ll m_e$, and $\lambda_{e^-}$ ($\lambda_{e^+}$) is the spin projection quantum number of $e^-$ ($e^+$) along the direction of $\vec{k}$ ($-\vec{k}$).
The amplitude for p-Ps to decay into $|k_1\rangle_{\gamma_1} |k_2\rangle_{\gamma_2}~(k_1,k_2=\pm 1)$  is given by
\begin{align}
\mathcal{M}_{k_1,k_2} & \approx \sqrt{\frac{1}{2m_e}} \psi(\vec{0}) \left(\mathcal{M}^\prime_{\lambda_{e^-}=\frac{1}{2},\lambda_{e^+}=\frac{1}{2};k_1,k_2}+\mathcal{M}^\prime_{\lambda_{e^-}=-\frac{1}{2},\lambda_{e^+}=-\frac{1}{2};k_1,k_2}\right)\label{m3}
\end{align}
with 
\begin{align}
\mathcal{M}^\prime_{\lambda_{e^-},\lambda_{e^+};k_1,k_2}=\lim_{\left|\vec{k}\right|\to 0} ~_{\gamma_1}\langle k_1|~_{\gamma_2}\langle k_2|
\left|\lambda_{e^-};\vec{k}\right\rangle_{e^-}\left|\lambda_{e^+};-\vec{k}\right\rangle_{e^+}~.
\end{align}
To obtain Eq. (\ref{m3}), we approximately treat $e^-$ and $e^+$ as static since $v=0$ ($|\vec{k}|=0$) at the LO level in the the non-relativistic limit of QED. And we also use the approximation $2m_e\approx m_\text{p-Ps}$. 
In Eq. (\ref{m3}), $\psi(\vec{0})$ represents the wave function of p-Ps at the origin ($\vec{0}$), obtained from the Fourier transformation of $\widetilde{\psi}(\vec{k})$ as
\begin{align}
\int \frac{d^3 \vec{k}}{(2 \pi)^3} \widetilde{\psi}(\vec{k})=\psi(\vec{0})~.
\end{align} 

Direct calculation yields the following amplitude relations of  
\begin{align}  
\mathcal{M}_{1,1}=-\mathcal{M}_{-1,-1}~,\quad \mathcal{M}_{1,-1}=\mathcal{M}_{-1,1}=0~. \label{aeq} 
\end{align}  
It is noteworthy that although the relations in Eq.~(\ref{aeq}) are derived at the LO level in the non-relativistic limit of QED, they remain exactly valid due to the constraints imposed by Clebsch-Gordan (CG) coefficients.  
For the $\gamma_1\gamma_2$ system produced through p-Ps decay without decoherence, the coefficients $\alpha_{k_1,k_2}$ in Eq.~(\ref{djxs}) are proportional to $\mathcal{M}_{k_1,k_2}$.
This gives
\begin{align}  
    \alpha_{1,1}=-\alpha_{-1,-1}=\frac{1}{\sqrt{2}},\quad \alpha_{1,-1}=\alpha_{-1,1}=0~,\label{aaaa}
\end{align}  
resulting in $\delta=-1$.
This explicitly demonstrates that the $\gamma_1\gamma_2$ system maintains quantum entanglement in the absence of decoherence, specifically forming a maximally entangled Bell state.  

\subsection{Joint Angular Distribution and Decoherence Effects}\label{sec:3-2} 

For the double Compton scattering of the $\gamma_1\gamma_2$ pair produced in p-Ps decay, under decoherence-free conditions, the joint angular distribution of the final-state photons $\gamma_1^\prime\gamma_2^\prime$ is given by
\begin{align}
    &\mathcal{W}(\theta_1,\theta_2,\phi_1,\phi_2)=\frac{5184\pi^2}{(40-27\ln 3)^2}\left(
    \frac{2+(1-\cos\theta_1)^3}{2-\cos\theta_1}
    \frac{2+(1+\cos\theta_2)^3}{2+\cos\theta_2}-\sin^2\theta_1 \sin^2\theta_2 \cos(2(\phi_1-\phi_2))
    \right)~,\label{pwe}
\end{align}
which is obtained by substituting the results of Eq.~(\ref{aaaa}) into Eq.~(\ref{ysfb}).
We note that Eq.~(\ref{pwe}), is closely related to the Pryce-Ward equation \cite{Pryce1947}. The apparent sign difference in front of \(\cos\theta_2\) arises from our choice of a common polar axis along the momentum of \(\gamma_1\), whereas Pryce and Ward choose the momentum of each photon as its own polar axis. For \(\gamma_2\), the scattering angle relative to its incident direction is \(\theta_2^{\rm PW}=\pi-\theta_2\), so that \(\cos\theta_2=-\cos\theta_2^{\rm PW}\). Substituting this into Eq.~(\ref{pwe}) yields the standard Pryce-Ward form, up to an overall normalization factor.

If decoherence occurs, the $\gamma_1\gamma_2$ system collapses into either the $|1\rangle_{\gamma_1} |1\rangle_{\gamma_2}$ or $|-1\rangle_{\gamma_1} |-1\rangle_{\gamma_2}$ states.
Assuming the fraction of decohered $\gamma_1\gamma_2$ pairs is $\rho$, the joint angular distribution becomes $\mathcal{W}^{\prime}(\rho;\theta_1,\theta_2,\phi_1,\phi_2)$, which can be expressed as
\begin{align}
   &\mathcal{W}^{\prime}(\rho;\theta_1,\theta_2,\phi_1,\phi_2) \nonumber \\
   =& \frac{5184\pi^2}{(40-27\ln 3)^2}\left(
    \frac{2+(1-\cos\theta_1)^3}{2-\cos\theta_1}
    \frac{2+(1+\cos\theta_2)^3}{2+\cos\theta_2}-(1-\rho)\sin^2\theta_1 \sin^2\theta_2 \cos(2(\phi_1-\phi_2))
    \right)~,\label{ffbb}
\end{align}
giving
\begin{align}
&\langle\mathcal{O}_1\rangle=\langle\mathcal{O}^\prime_1(\theta_1,\theta_2)\rangle=-(1-\rho)~.
\end{align}
Therefore, in practical measurements, the fraction of decohered $\gamma_1\gamma_2$ pairs can be quantified from the observed values of $\langle\mathcal{O}_1\rangle$ and $\langle\mathcal{O}^\prime_1(\theta_1,\theta_2)\rangle$.  

\section{Feasibility of a Local Hidden-Variable Description} \label{sec:4} 

This section addresses the central question of whether the quantum correlations predicted in Section \ref{sec:3} can be accommodated within a framework of local realism. As a foundation for this analysis, we first establish the symmetry properties of single-photon Compton scattering within QFT. Then, we try to construct an LHVT designed to replicate the joint angular distribution $\mathcal{W}^{\prime}(\rho;\theta_1,\theta_2,\phi_1,\phi_2)$ in Eq.~(\ref{ffbb}), taking angular-momentum conservation, which is a fundamental constraint from the spin-0 nature of p-Ps, and the rotation invariance as the guiding principles for the hidden-variable ansatz. The analysis then pivots on examining the physical constraints that can be imposed on such a theory. We demonstrate that the demand of mirror symmetry with respect to the plane spanned by the photon spin and momentum vectors serves as a decisive, physically motivated criterion.  

\subsection{Symmetry Analysis of Single-Photon Compton Scattering in QFT} \label{sec:4-0} 

In this subsection, we analyze single-photon Compton scattering under QFT, determining how the angular distribution of the final-state photon depends on the spatial orientation of the incident photon spin, and we examine the symmetries satisfied by the angular distribution of the final-state photon. For convenience, we continue to denote the process as $\gamma_1+e^-_1 \to \gamma_1^\prime+e^{-\prime}_1$, and the parametrization of the four-momenta of all particles still follows Eqs. (\ref{rsg}) to (\ref{csg}).

We denote the three-momentum direction of $\gamma_1$ by $\hat{z}=(0, 0, 1)$. We define the unit direction vector $\hat{n}$ as $(\sin x \cos y, \sin x \sin y, \cos x)$.
so that
\begin{align}
    & \hat{R}(y,x,y^\prime)\hat{z}=\hat{n}~.
\end{align}
Here, $\hat{R}(y,x,y^\prime)$ denotes a spatial rotation specified by the Euler angles $y$, $x$, and $y^\prime$.
Assume that the spin of $\gamma_1$ points along $\hat{n}$; that is, its spin projection along $\hat{n}$ equals $1$, corresponding to the polarization state $\left|1\right\rangle_{\hat{n}}$. The relation between $\left|1\right\rangle_{\hat{n}}$ and the polarization state with spin projection $1$ along $\hat{z}$, $\left|1\right\rangle_{\hat{z}}$, is
\begin{align}
   & \left|1\right\rangle_{\hat{n}}=\hat{R}(y,x,y^\prime) \left|1\right\rangle_{\hat{z}}=\sum_{k=-1}^1 D^1_{1,k}(y,x,y^\prime)\left|k\right\rangle_{\hat{z}}
\end{align}
with
\begin{align}
&D^1_{1,k}(y,x,y^\prime)=e^{-iky-iy^\prime}d^1_{1,k}(x)~,\\
&d^1_{1,\pm1}(x)=\frac{1\pm\cos x}{2}~, \quad d^1_{1,0}(x)=-\frac{\sin x}{\sqrt{2}}~.
\end{align}
Since photons are massless, the polarization state $\left|0\right\rangle_{\hat{z}}$ does not exist. Dropping the $\left|0\right\rangle_{\hat{z}}$ component in $\left|1\right\rangle_{\hat{n}}$ and renormalizing, we obtain
\begin{align}
\left|1\right\rangle_{\hat{n}}&=\frac{1}{\sqrt{2(1+\cos^2 x)}}\left(e^{-iy-iy^\prime}(1+\cos x)\left|1\right\rangle_{\hat{z}}+e^{iy-iy^\prime}(1-\cos x)\left|-1\right\rangle_{\hat{z}}\right)\\
&=\sum_{k=\pm 1}u_k \left|k\right\rangle_{\hat{z}}~.
\end{align}
So, for incident photon $\gamma_1$ prepared in the polarization state $\left|1\right\rangle_{\hat{n}}$, the angular distribution of the final-state photon $\gamma_1^\prime$ is
\begin{align}  
   &\mathcal{F}(x,y;\theta_1,\phi_1)=\frac{1}{\sqrt{\Gamma}}\frac{1}{16\pi^2}q_1^2 \sum_{\substack{j_1,n_1 = \pm \frac{1}{2} \\ m_1=\pm1}} \left|\sum_{k=\pm 1}  u_{k}\mathcal{M}^1_{k;j_1,m_1,n_1}\right|^2  \nonumber\\
   =& \frac{9}{2\pi (40-27\ln 3)}\left(\frac{2+(1-\cos\theta_1)^3}{(2-\cos\theta_1)^3}+\frac{\sin^2 x}{1+\cos^2 x}\frac{\sin^2\theta_1}{(2-\cos\theta_1)^2}\cos(2(\phi_1-y))\right)~,
\end{align}
which satisfies
\begin{align}  
   &\int_{-1}^1d\cos\theta_1\int_0^{2\pi}d\phi_1~ \mathcal{F}(x,y;\theta_1,\phi_1)=1~.
\end{align}

We now discuss the transformation invariance of \(\mathcal{F}(x,y;\theta_1,\phi_1)\). It is instructive to first consider the special case \(y=0\). In this case the spin direction \(\hat{n}\) lies in the \(xz\)-plane, and the plane spanned by the photon momentum and \(\hat{n}\) coincides with the \(xz\)-plane. Reflection about this plane reverses the sign of the azimuthal angle \(\phi_1\), and the angular distribution is invariant, i.e., \(\mathcal{F}(x,0;\theta_1,\phi_1)=\mathcal{F}(x,0;\theta_1,-\phi_1)\). This is the mirror symmetry with respect to the plane spanned by the photon momentum and \(\hat{n}\). More generally, for arbitrary \(y\), the mirror plane is obtained by rotating the \(xz\)-plane about the \(z\)-axis by \(y\), which leads to the corresponding azimuthal reflection. With this intuitive picture in mind, we now turn to a more precise and complete treatment. It can be shown that \(\mathcal{F}(x,y;\theta_1,\phi_1)\) is invariant under the following transformations:
\begin{itemize}
    \item $x\to x^\prime=\pi-x$. 
    This implies that the angular distribution of the final-state photon induced by the spin orientation $(x,y)$ is identical to that induced by $(\pi-x,y)$.
    \item $\phi_1\to \phi_1^\prime= d\pi +y-(\phi_1-y)$, where $d=0$ or $1$. 
    Since the direction specified by $(\theta_1,\phi_1)$ and that specified by $(\theta_1,\phi_1^\prime)$ with $d=0$ are related by a reflection with respect to the plane defined by $\hat{z}$ and $\hat{n}$, $\mathcal{F}(x,y;\theta_1,\phi_1)$ exhibits mirror symmetry.
\end{itemize}
Moreover, parity invariance of the QED process requires
\begin{align}
    \mathcal{F}(x,y;\theta_1,\phi_1)=\mathcal{F}(\pi-x,-y;\theta_1,\pi-\phi_1)~,
\end{align}
which is already ensured by the invariance of $\mathcal{F}(x,y;\theta_1,\phi_1)$ under the transformations $x\to x^\prime$ and $\phi_1\to \phi_1^\prime$ with $d=1$.

\subsection{Hidden-Variable Ansatz Based on Angular-Momentum Conservation} \label{sec:4-1} 

Since p-Ps has spin 0, angular-momentum conservation implies that the spin angular momenta of the decay photons $\gamma_i~(i=1,2)$, denoted $\vec{S}_i$, satisfy $\vec{S}_1=-\vec{S}_2$.
We denote the unit direction vectors of $\vec{S}_i$ by $\hat{S}_i$ and parameterize them as
\begin{align}
     &\hat{S}_1=-\hat{S}_2=\hat{S}=(\sin x \cos y,\sin x \sin y,\cos x)~.
\end{align}
Here, $x$ and $y$ serve as the local hidden variables. We denote their joint probability distribution by $G(x,y)$, which satisfies
\begin{align}
     & G(x,y)\ge 0~, \quad \int_{-1}^1 d\cos x\int_0^{2\pi}dy~G(x,y)=1~.
\end{align}

For $y=0$ and a given $x$, we denote the joint distribution of $\theta_1$ and $\phi_1$ by $F(x;\theta_1,\phi_1)$, and we require this function to satisfy
\begin{align}
     &F(x;\theta,\phi)\ge0~, \quad \int_{-1}^1 d\cos\theta \int_0^{2\pi}d\phi~F(x;\theta,\phi)=1~.
\end{align}
For given $x$ and $y$, we denote the joint angular distribution of $\theta_i$ and $\phi_i$ by $F_i(x,y;\theta_i,\phi_i)$. Then, by rotational invariance, we have
\begin{align}
     & F_1(x,y;\theta_1,\phi_1)=F(x;\theta_1,\phi_1-y)~, \\
     & F_2(x,y;\theta_2,\phi_2)=F(x;\pi-\theta_2,\pi+y-\phi_2)~.
\end{align}
Thus, within the local hidden-variable framework, $\mathcal{W}^{\prime}(\rho;\theta_1,\theta_2,\phi_1,\phi_2)$ satisfies
\begin{align}
    &\frac{1}{256\pi^4}q_1^2 q_2^2 \mathcal{W}^{\prime}(\rho;\theta_1,\theta_2,\phi_1,\phi_2) \nonumber\\
    =& \int_{-1}^1d\cos x  \int_0^{2\pi}dy~G(x,y)~F(x;\theta_1,\phi_1-y)~F(x;\pi-\theta_2,\pi+y-\phi_2)~. \label{pzz}
\end{align}

\subsection{Mirror Symmetry as a Criterion for the Constructibility} \label{sec:4-2} 

We expand $F(x;\theta,\phi)$ in spherical harmonics as
\begin{align}
     &F(x;\theta,\phi)=\frac{1}{4\pi}+\sum_{l=1}^{\infty}\sum_{m=-l}^{l}a_{l,m}(x)Y_{l,m}(\theta,\phi)~.
\end{align}
Here, $a_{0,0}(x)Y_{0,0}(\theta,\phi)$ is fixed as $\frac{1}{4\pi}$ due to the normalization condition of $F(x;\theta,\phi)$.
Since $F(x;\theta,\phi)$ is real-valued, it follows that
\begin{align}
     & a_{l,m}(x)=(-1)^m a^*_{l,-m}(x)~.
\end{align}
We choose
\begin{align}
h=\sin^2\theta_1 \sin^2\theta_2 \cos(2\phi_1-2\phi_2)~.
\end{align}
Since
\begin{align}
&h=\frac{16\pi}{15}\left(Y^*_{2,-2}(\theta_1,\phi_1-y)Y^*_{2,-2}(\pi-\theta_2,y+\pi-\phi_2)+Y^*_{2,2}(\theta_1,\phi_1-y)Y^*_{2,2}(\pi-\theta_2,y+\pi-\phi_2)\right)~,
\end{align}
within LHVT the right-hand side of Eq.~(\ref{pzz}) yields
\begin{align}
& \langle h\rangle_{\text{LHVT}}= \frac{16\pi}{15}\int_{-1}^1d\cos x  \int_0^{2\pi}dy~G(x,y)\left(a^2_{2,-2}(x)+a^2_{2,2}(x)\right)~.\label{zjx}
\end{align}
We have used the orthogonality of spherical harmonics in obtaining Eq.~(\ref{zjx}).
In QFT, the left-hand side of Eq.~(\ref{pzz}) yields
\begin{align}
& \langle h\rangle_{\text{QFT}}= -288(1-\rho)\left(\frac{10-9\ln 3}{40-27\ln 3}\right)^2~.
\end{align}

The mirror symmetry analyzed in Section~\ref{sec:4-0} is not merely a property of the QFT scattering amplitude but can be elevated to a physical principle. We now examine its role in constructing an LHVT to replicate the joint angular distribution $\mathcal{W}^{\prime}(\rho;\theta_1,\theta_2,\phi_1,\phi_2)$.
Let $T$ denote the plane spanned by the two vectors, $\hat{z}$ and $\hat{S}$. Depending on whether the distribution function $F_i(x,y;\theta_i,\phi_i)$ is symmetric with respect to the plane $T$, we consider the following cases:
\begin{itemize}
    \item When the distribution functions $F_i(x,y;\theta_i,\phi_i)$ are symmetric with respect to the plane $T$, we have $F(x;\theta,\phi)=F(x;\theta,-\phi)$.
This further requires that $a_{l,m}(x)$ be real-valued.
In this case, $\langle h\rangle_{\text{LHVT}}\ge 0$.
By contrast, for $0\le\rho<1$, $\langle h\rangle_{\text{QFT}}<0$.
Since $\langle h\rangle_{\text{LHVT}}$ and $\langle h\rangle_{\text{QFT}}$ cannot be equal, an LHVT cannot reproduce $\mathcal{W}^{\prime}(\rho;\theta_1,\theta_2,\phi_1,\phi_2)$ as long as the entangled photon pairs have not completely decohered ($0\le\rho<1$).
\item When no symmetry with respect to the plane $T$ is imposed on the distribution functions $F_i(x,y;\theta_i,\phi_i)$, we find a set of solutions to Eq.~(\ref{pzz}):
\begin{align}
    &G(x,y)=\frac{1}{4\pi}~,\\
    & F(x;\theta,\phi)=\frac{9}{2\pi\left(40-27\ln 3\right)}\left(
    \frac{2+(1-\cos\theta)^3}{(2-\cos\theta)^3}+\sqrt{2(1-\rho)}\frac{\sin^2\theta}{(2-\cos\theta)^2}\cos\left(2\phi+\frac{\pi}{2}\right)
    \right)~.\label{flp}
\end{align}
Clearly, $F(x;\theta,\phi)$ is not identically equal to $F(x;\theta,-\phi)$. 
Since
\begin{align}
    & \frac{2+(1-\cos\theta)^3}{(2-\cos\theta)^3}>\sqrt{2}\frac{\sin^2\theta}{(2-\cos\theta)^2}~\text{for $\theta\in[0,\pi]$}~,
\end{align}
the non-negativity of $F(x;\theta,\phi)$ is guaranteed.
It is noteworthy that Bell’s LHVT construction given in Appendix I of Ref.~\cite{Kasday1971} can be obtained from the solution in Eq.~(\ref{flp}) via the substitution\footnote{By comparing the definitions, we identify $\theta_1=\Theta_1$, $\theta_2=\pi-\Theta_2$, and $\phi_{1,2}=\Phi_{1,2}$.}:
\begin{align}
    y=\lambda+\frac{\pi}{4}~.
\end{align}
We emphasize that the solutions are not unique.
So, in the case that no symmetry with respect to the plane $T$ is imposed on the distribution functions $F_i(x,y;\theta_i,\phi_i)$, there exist LHVTs that reproduce $\mathcal{W}^{\prime}(\rho;\theta_1,\theta_2,\phi_1,\phi_2)$.
\end{itemize}

Therefore, under the indisputable fundamental principles of angular-momentum conservation and the rotation invariance, a measurement of $\langle h\rangle$ with sufficient precision, combined with an independent verification of mirror symmetry in Compton scattering, would rule out any LHVT that respects this symmetry.

\section{Conclusion} \label{sec:con} 

In this work, we have proposed and analyzed a scheme to probe photon entanglement originating from p-Ps decay using Compton scattering azimuthal correlations. The correlation observable $\mathcal{O}_1$ provides a clear signature of entanglement and a quantitative measure of decoherence.

Our primary theoretical finding addresses the feasibility of a local hidden-variable description. Our analysis started by establishing, within QFT, the mirror symmetry inherent to the Compton scattering of a single photon. We showed that an LHVT based on angular-momentum conservation and the rotation invariance can, in principle, reproduce the measured joint angular distribution $\mathcal{W}^{\prime}(\rho;\theta_1,\theta_2,\phi_1,\phi_2)$, provided no specific symmetry constraints are imposed on the underlying distribution $F(x;\theta,\phi)$. However, when the physically motivated mirror symmetry condition is enforced, the LHVT prediction for a key correlation term becomes strictly non-negative. This stands in direct contradiction to the negative value predicted by QFT for any partially entangled state ($0 \le \rho < 1$). Consequently, the demand for mirror symmetry within the LHVT framework acts as a powerful criterion: it precludes a local realistic description of the entangled photon pair correlations unless the state has completely lost its quantum coherence.

Our work does not claim to close the LHVT loophole in all possible models, but rather shows that symmetry principles can serve as novel, physically motivated constraints in foundational tests.
This result underscores that the failure of local realism can manifest not only through violations of Bell inequalities but also through the incompatibility of observed correlations with natural symmetry constraints imposed on putative hidden-variable models.
Our analysis, centered on a fundamental QED process like p-Ps decay, suggests that Compton scattering of entangled photons, coupled with symmetry-based reasoning, offers a viable and alternative pathway for testing the foundations of quantum mechanics. Future experimental measurements of the azimuthal correlations studied here could directly confront the mirror-symmetry-restricted LHVT predictions, providing a fresh testbed for local realism.

\begin{acknowledgments}
J. Pei is supported by the National Natural Science Foundation of China (Project No. 12505121), by the Joint Fund of Henan Province Science and Technology R$\&$D Program (Project No. 245200810077), by the Startup Research Fund of Henan Academy of Sciences (Project No. 20251820001),
and by the Scientific and Technological Research Project of Henan Academy of Sciences (Project No. 20262320001).
L Wu is supported by the National Natural Science Foundation of China (Project No. 12605118), and by the Natural Science Basic Research Program of Shaanxi (Project No. 2024JC YBMS-039).
\end{acknowledgments}

\appendix

\section{Expressions of $\tilde{\mathcal{M}}^i_{k;j,m,n}$}\label{appen}

With $q_i$ given in Eq.~(\ref{q12}), we define
\begin{align}
    & p=\sqrt{\left(2-q_1\right)^2-1}~, \\
    & z_1=\sqrt{2-p-q_1}~,\\
    & z_2=\sqrt{2+p-q_1}~,\\
    & z_3=\sqrt{p(1+p-q_1\cos\theta_1)}~, \\
    & b=\sqrt{\left(2-q_2\right)^2-1}~, \\
    & y_1=\sqrt{2-b-q_2}~,\\
    & y_2=\sqrt{2+b-q_2}~,\\
    & y_3=\sqrt{b(-1+b-q_2\cos\theta_2)}~.
\end{align}
Then, $\tilde{\mathcal{M}}^i_{k;j,m,n}$ are expressed as
\begin{align}
  & \tilde{\mathcal{M}}^1_{-1;-\frac{1}{2},-1,-\frac{1}{2}}= \frac{z_1 z_3 \left(1+\cos \theta_1\right)}{\sqrt{2} p} ~,\\
  & \tilde{\mathcal{M}}^1_{1;-\frac{1}{2},-1,-\frac{1}{2}}=\frac{z_1 z_3 \left(1-\cos \theta_1\right)}{\sqrt{2} p} ~,\\
  & \tilde{\mathcal{M}}^1_{-1;-\frac{1}{2},-1,\frac{1}{2}}=-\frac{q_1 z_2 \sin \theta_1 \left(1+\cos \theta_1\right)}{\sqrt{2} z_3} ~,\\
  & \tilde{\mathcal{M}}^1_{1;-\frac{1}{2},-1,\frac{1}{2}}=-\frac{q_1 z_2 \sin \theta_1 \left(1-\cos \theta_1\right)}{\sqrt{2} z_3} ~,\\
  & \tilde{\mathcal{M}}^1_{-1;-\frac{1}{2},1,-\frac{1}{2}}=\frac{z_2 z_3 \left(1-\cos \theta_1\right)}{\sqrt{2} p} ~,\\
  & \tilde{\mathcal{M}}^1_{1;-\frac{1}{2},1,-\frac{1}{2}}=\frac{ \left(1+\cos \theta_1\right) (p-q_1+1)z_1+ \left(1-\cos ^2\theta_1\right)q_1 z_2}{\sqrt{2} z_3} ~,\\
  & \tilde{\mathcal{M}}^1_{-1;-\frac{1}{2},1,\frac{1}{2}}=-\frac{q_1 z_1 \sin \theta_1 \left(1-\cos \theta_1\right)}{\sqrt{2} z_3} ~,\\
  & \tilde{\mathcal{M}}^1_{1;-\frac{1}{2},1,\frac{1}{2}}=- \frac{\sin \theta_1 \left( \left(-p+q_1 \cos \theta_1-1\right)z_1+ (p+q_1+1)z_2\right)}{\sqrt{2} z_3} ~,\\
  & \tilde{\mathcal{M}}^1_{-1;\frac{1}{2},-1,-\frac{1}{2}}= \frac{\sin \theta_1 \left( \left(-p+q_1 \cos \theta_1-1\right)z_1+ (p+q_1+1)z_2\right)}{\sqrt{2} z_3} ~,\\
  & \tilde{\mathcal{M}}^1_{1;\frac{1}{2},-1,-\frac{1}{2}}=\frac{q_1 z_1 \sin \theta_1 \left(1-\cos \theta_1\right)}{\sqrt{2} z_3} ~,\\
  & \tilde{\mathcal{M}}^1_{-1;\frac{1}{2},-1,\frac{1}{2}}=\frac{ \left(1+\cos \theta_1\right) (p-q_1+1)z_1+ \left(1-\cos ^2\theta_1\right)q_1 z_2}{\sqrt{2} z_3} ~,\\
  & \tilde{\mathcal{M}}^1_{1;\frac{1}{2},-1,\frac{1}{2}}=\frac{z_2 z_3 \left(1-\cos \theta_1\right)}{\sqrt{2} p} ~,\\
  & \tilde{\mathcal{M}}^1_{-1;\frac{1}{2},1,-\frac{1}{2}}= \frac{q_1 z_2 \sin \theta_1 \left(1-\cos \theta_1\right)}{\sqrt{2} z_3} ~,\\
  & \tilde{\mathcal{M}}^1_{1;\frac{1}{2},1,-\frac{1}{2}}=\frac{q_1 z_2 \sin \theta_1 \left(1+\cos \theta_1\right)}{\sqrt{2} z_3} ~,\\
  & \tilde{\mathcal{M}}^1_{-1;\frac{1}{2},1,\frac{1}{2}}=\frac{z_1 z_3 \left(1-\cos \theta_1\right)}{\sqrt{2} p} ~,\\
  & \tilde{\mathcal{M}}^1_{1;\frac{1}{2},1,\frac{1}{2}}=\frac{z_1 z_3 \left(1+\cos \theta_1\right)}{\sqrt{2} p}~, \\
   & \tilde{\mathcal{M}}^2_{-1;-1,-1,-1}= \frac{ \left(1-\cos \theta _2\right) (-b-q_2+1)y_2+ \left(1-\cos ^2\theta _2\right)q_2 y_1}{\sqrt{2} y_3} ~,\\
   & \tilde{\mathcal{M}}^2_{1;-1,-1,-1}=-\frac{y_1 y_3 \left(1+\cos \theta _2\right)}{\sqrt{2} b} ~,\\
   & \tilde{\mathcal{M}}^2_{-1;-1,-1,1}=-\frac{\sin\theta _2 \left( \left(-b+q_2 \cos \theta _2+1\right)y_2+ (b-q_2-1)y_1\right)}{\sqrt{2} y_3} ~,\\
   & \tilde{\mathcal{M}}^2_{1;-1,-1,1}=\frac{q_2 y_2 \sin\theta _2 \left(1+\cos \theta _2\right)}{\sqrt{2} y_3} ~,\\
   & \tilde{\mathcal{M}}^2_{-1;-1,1,-1}=-\frac{y_2 y_3 \left(1+\cos \theta _2\right)}{\sqrt{2} b} ~,\\
   & \tilde{\mathcal{M}}^2_{1;-1,1,-1}=-\frac{y_2 y_3 \left(1-\cos \theta _2\right)}{\sqrt{2} b} ~,\\
   & \tilde{\mathcal{M}}^2_{-1;-1,1,1}=\frac{q_2 y_1 \sin\theta _2 \left(1+\cos \theta _2\right)}{\sqrt{2} y_3} ~,\\
   & \tilde{\mathcal{M}}^2_{1;-1,1,1}=\frac{q_2 y_1 \sin\theta _2 \left(1-\cos \theta _2\right)}{\sqrt{2} y_3} ~,\\
   & \tilde{\mathcal{M}}^2_{-1;1,-1,-1}=-\frac{q_2 y_1 \sin\theta _2 \left(1-\cos \theta _2\right)}{\sqrt{2} y_3} ~,\\
   & \tilde{\mathcal{M}}^2_{1;1,-1,-1}=-\frac{q_2 y_1 \sin\theta _2 \left(1+\cos \theta _2\right)}{\sqrt{2} y_3} ~,\\
   & \tilde{\mathcal{M}}^2_{-1;1,-1,1}=-\frac{y_2 y_3 \left(1-\cos \theta _2\right)}{\sqrt{2} b} ~,\\
   & \tilde{\mathcal{M}}^2_{1;1,-1,1}=-\frac{y_2 y_3 \left(1+\cos \theta _2\right)}{\sqrt{2} b} ~,\\
   & \tilde{\mathcal{M}}^2_{-1;1,1,-1}=-\frac{q_2 y_2 \sin\theta _2 \left(1+\cos \theta _2\right)}{\sqrt{2} y_3} ~,\\
   & \tilde{\mathcal{M}}^2_{1;1,1,-1}=\frac{\sin\theta _2 \left(\left(-b+q_2 \cos \theta _2+1\right)y_2 + (b-q_2-1)y_1\right)}{\sqrt{2} y_3} ~,\\
   & \tilde{\mathcal{M}}^2_{-1;1,1,1}=-\frac{y_1 y_3 \left(1+\cos \theta _2\right)}{\sqrt{2} b} ~,\\
   & \tilde{\mathcal{M}}^2_{1;1,1,1}=\frac{  \left(1-\cos \theta _2\right)(-b-q_2+1)y_2+ \left(1-\cos ^2\theta _2\right)q_2 y_1}{\sqrt{2} y_3}~.
\end{align}

\nocite{*}

\bibliography{prd}

\end{document}